\documentclass[pra,showpacs,twocolumn,showkeys]{revtex4-1}
\usepackage{graphics}
\usepackage{amsmath}
\usepackage{color}
\begin{document}

\title{The effect of extreme confinement on the nonlinear-optical response of quantum wires}
\author{Shoresh Shafei}
\author{Mark G. Kuzyk}
\email{kuz@wsu.edu}
\affiliation{Department of Physics and Astronomy, Washington State University, Pullman, Washington  99164-2814}

\begin{abstract}
This work focuses on understanding the nonlinear-optical response of a 1-D quantum wire embedded in 2-D space when quantum-size effects in the transverse direction are minimized using an extremely weighted delta function potential.  Our aim is to establish the fundamental basis for understanding the effect of geometry on the nonlinear-optical response of quantum loops that are formed into a network of quantum wires. Using the concept of \emph{leaky} quantum wires, it is shown that in the limit of full confinement, the sum rules are obeyed when the transverse infinite-energy continuum states are included.  While the continuum states associated with the transverse wavefunction do not contribute to the nonlinear optical response, they are essential to preserving the validity of the sum rules. This work is a building block for future studies of nonlinear-optical enhancement of quantum graphs (which include loops and bent wires) based on their geometry. These properties are important in quantum mechanical modeling of any response function of quantum-confined systems, including the nonlinear-optical response of any system in which there is confinement in at leat one dimension, such as nanowires, which provide confinement in two dimensions.
\end{abstract}


\maketitle
\vspace{1em}
\section{Introduction}
The study of nonlinear optical (NLO) properties of materials is motivated by both the beauty of the underlying physics of light-matter interactions as well as its potential for applications such as 3-D nanophotolithography \cite{kawat01.01}, telecommunications \cite{wang04.01}, and designing new materials \cite{karot04.01} for cancer therapies \cite{roy03.01}, to name a few.  Quantum theories of the nonlinear optical response have deepened our understanding of the basic science that has led to the identification of general characteristics of systems for which the nonlinear response is optimized,\cite{marde91.01,meyer94.01,zhou06.01,zhou07.02} some of which have been experimentally verified.\cite{marde93.01,risse03.01,rumi00.01,perez07.01}

The fundamental limits of the first\cite{kuzyk00.01} and second\cite{kuzyk00.02} hyperpolarizabilities in the off-resonant regime depend on the number of electrons in the atom or molecule, $N$, and the excitation energy from the ground to the first excited state, $E_{10}$. In calculating these limits, the generalized Thomas-Reiche-Kuhn sum rules are used to simplify the sum-over-states expressions of Orr and Ward,\cite{orr71.01} followed by the application of the Three Level Ansatz, which has not been proven mathematically but has been confirmed to hold in every case studied.\cite{kuzyk05.01,kuzyk09.01} The three-level ansatz, which states that when the hyperpolarizability of a quantum system is at the fundamental limit only three states contribute,\cite{kuzyk09.01} has also been used to calculate the fundamental limit of first hyperpolarizability in the \emph{resonant} regime.\cite{kuzyk06.03}

A tabulation of experimental results reveals that the largest measured hyperpolarizabilities are smaller than the fundamental limit by factor of 30.\cite{kuzyk03.01,kuzyk03.02}  This gap is attributed to the unfavorable distribution of the energy eigenstates.\cite{Tripa04.01,perez07.02} The criticality of the energy-level spacing has been confirmed in Monte Carlo studies.\cite{shafe11.01}

Despite the fact that methods such as modulated conjugation have been recently proposed to yield molecules with enhanced NLO response,\cite{perez07.01,perez09.01} it is difficult to ``fabricate" molecules with these desired properties. Thus it is reasonable to investigate new approaches and/or material classes such as quantum-confined systems (QCSs), which include multiple quantum wells\cite{cheml89.01,agran93.04,hall2010using} and quantum wires \cite{guble99.01,liebe07.01,tian.09.01,wang10.01,yan09.01,duan2001indium}.

QCSs have been extensively studied both theoretically and experimentally as tiny labs to investigate quantum mechanical effects.\cite{ashoori1996electrons} The scaling of the NLO response of QCSs provides the means for making materials with controllable NLO properties with applications in optical devices such as lasers, solar cells, and nonlinear-optical switches.\cite{banfi1998nonlinear} Indeed, semiconductor nanowires are being considered as building blocks of optical devices. Such nanowires are structures with diameters of 1-100 nm and lengths of several micrometers. During nanowire synthesis, key parameters such as chemical composition, diameter, and length can be controlled, enabling a wide range of devices and applications such as $p-n$ diodes, LED's, transistors and nano-scale lasers.\cite{liebe07.01, yan09.01} In nonlinear optics, photonic nanowires have found applications in the generation of single-cycle pulses and optical processing with sub-mW powers.\cite{foster2008nonlinear}

In early studies, Hache \emph{et al.} reported on the effects of quantum confinement on the NLO properties of metal colloids inside a dielectric near the surface plasma resonance.\cite{hache86.02} They showed that the large enhancement of the third order nonlinear response originates in the electrons that are confined to the spherical metal particles, with the response scaling as $a^{-3}$, where $a$ is the radius of the sphere. In other work, Chen \emph{et al.} showed that exciton localization due to confinement effects are mainly responsible for the large third order NLO enhancement in Silicon nanowires,\cite{chen1993enhancement} scaling as $a_0^{-6}$, where $a_0$ is the Bohr radius of the exciton.  The goal of these studies was to understand how quantum-confinement can be used to increase the NLO response of QCSs, including 1-D confined quantum wells, 2-D confined quantum wires and 3-D confined quantum dots.

The focus of the present work is to lay the foundations for studying the effects of the geometry of quantum graphs made of networks of quantum wires on their NLO properties. For this purpose, we study in detail a single 1-D wire embedded in 2-D space and show that in the limit of extreme transverse quantum confinement, only the longitudinal component contributes to the NLO response. By minimizing the role of quantum-confinement in the NLO response, the effects of geometry on NLO enhancement can be unambiguously determined.  We also reconcile the paradox of how the transverse sum rules can be satisfied when the contribution of the transverse states to the nonlinear response vanishes.

\section{System of quantum wires}

\begin{figure}
  \includegraphics{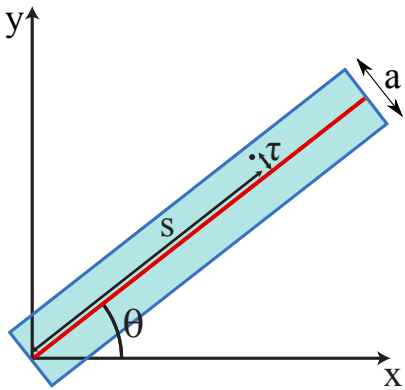}
  \caption{A single quantum wire in 2-D space. The red line shows the wire, which is defined by the transverse delta function potential. The blue region is a box to which the electron is confined. The width of the box, $a$, is made arbitrarily large to represent the continuum states. $s$ and $\tau$ are longitudinal and transverse coordinates of the electron.}\label{fig:2Dwire}
\end{figure}

Electrons move freely along a quantum wire, which we define as the longitudinal direction, $\hat s$, and are tightly confined in the transverse direction, $\hat\tau $, as illustrated in Fig. \ref{fig:2Dwire}. For simplicity, we consider a single electron, which will behave as a free particle along $\hat s$ with energy eigenfunction,
\begin{equation}\label{wavefunction}
\psi(s) = A e^{i k_s s} + B e^{- i k_s s},
\end{equation}
where $A$, $B$, and $k_s$ are obtained from the boundary conditions and normalization. For a wire segment that is part of a larger structure, such as quantum graphs -- where several wire segments can meet at a node, continuity of probability density through each node must be used.\cite{harrison2005quantum}

We note that the single electron calculation can be generalized to metals and semiconductors by adding to the band structure the boundary conditions.  By determining the ground state configuration from the Fermi energy, the states of the system can be built up from the single electron states.  We therefore focus our our work solely on the one-electron calculation.

In the transverse direction, we use a strongly weighted delta-function potential to model confinement.  This is similar to work by Scott, who presented a solution to a particle in a delta potential inside a box.\cite{scott1963delta} A series of related papers include those of Damert, who verified the completeness of delta potential eigenstates,\cite{Damert75} Blinder, who calculated the Green's function and propagator for one dimensional delta function,\cite{blind88} Lapidus, who employed perturbation theory for a delta potential inside a box,\cite{lapidus1987particle} and Joglekar, who studied the delta potential in the weak and strong coupling limits.\cite{joglek09} Delta function potentials are also used to describe real systems such as one dimensional diatomic ions,\cite{lapidus1970one} and to model the electronic structure of graphene layers and nanotubes.\cite{hsu2005modeling}

The transverse motion of an electron in a thin wire can be described using a delta function potential of the form,
\begin{equation}\label{DeltaFunctionPotential}
V(\tau) = -g \delta(\tau) ,
\end{equation}
where $g>0$ is the strength of confinement. The transverse Hamiltonian of the system is given by
\begin{equation}\label{Htau}
H_\tau = -\frac{\hbar^2}{2 m}\partial^2_\tau- g\delta\left(\tau \right),
\end{equation}
where $m$ is the electron mass and $\partial_x \equiv \partial/\partial x$, leading to the Schrodinger Equation,
\begin{equation}\label{TranasversalEq}
\partial^2_\tau \eta(\tau) \pm k_\tau^2 \eta(\tau) = 0,
\end{equation}
and
\begin{equation}\label{k}
k_s^2 + k_\tau^2 = \frac{2mE}{\hbar^2}.
\end{equation}
The bound state solution of Eq. (\ref{TranasversalEq}) is given by,
\begin{equation}\label{TransverseGround}
\eta_0 (\tau) = \sqrt{k_0^\tau}e ^ {- k_0^\tau \left| \tau \right| } ,
\end{equation}
with transverse energy
\begin{equation}\label{energy}
E_0^\tau = - \frac {\hbar^2 {k_0^\tau}^2} {2 m},
\end{equation}
where
\begin{equation}\label{definektau0}
k_0^\tau = \frac {m g}{\hbar^2} .
\end{equation}
The ground state is the only bound state of the transverse wavefunction.

\begin{figure}
\includegraphics{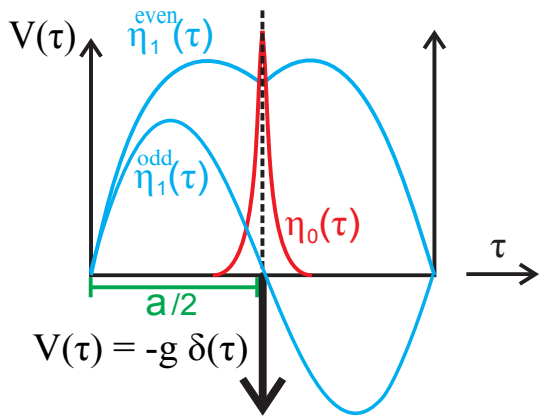}
\caption{The delta function in a box potential.  Shown are the ground state, $\eta_0 (\tau)$, the first even-parity continuum state, $\eta_1^{even}(\tau)$, and the first odd-parity continuum state, $\eta_1^{odd}(\tau)$.}
\label{fig:potential}
\end{figure}

For states with positive energies, which will represent the continuum states, we impose box normalization as shown by the potential energy function in Figure \ref{fig:potential} with width $a>>1/k_0^{\tau}$. The solution of the transverse wavefunction, Eq. (\ref{TranasversalEq}), for positive energies is,
\begin{eqnarray}\label{EnergyWavefunctions}
\eta(\tau) &=& \frac {\sqrt{2/a}} {\left(1 - \frac {\sin \left( k_\tau a \right)} {k_\tau a} \right)^{1/2}} \sin \left[k_\tau \left( \frac {a} {2} - |\tau|\right)\right] , \nonumber \\
\end{eqnarray}
where the discontinuity condition of the derivative of the transverse wavefunction, $\frac {d \eta} {d\tau}$, at $\tau = 0$, yields the transcendental equation
\begin{equation}\label{trans}
\tan \left( \frac {k_\tau a} {2} \right) = \frac {k_\tau} {k_0^\tau},
\end{equation}
from which $k_\tau$ is determined. The positive energy eigenvalues are given by
\begin{equation}\label{energy}
E_\nu = \frac {\hbar^2 {k_\nu^\tau}^2} {2m},
\end{equation}
where $k_\nu$ is the $\nu^{th}$ zero of Eq. (\ref{trans}).

At the limit of extreme confinement, when $g \rightarrow \infty$, Eq. (\ref{trans}) becomes,
\begin{equation}\label{transLimit}
\tan \left( \frac {k_\tau a} {2} \right) = 0,
\end{equation}
which yields,
\begin{equation}\label{EnergyEigenvalues}
\frac {k_\nu^\tau a} {2} = \nu \pi.
\end{equation}
In this case, the odd-parity excited states remain unchanged, the even parity excites states vanish at the origin and take the form $\eta_i^{even}(\tau) = \left| \eta_i^{odd}(\tau) \right| $ and the ground state wavefunction will be sharp and symmetric with exponential tails.

Using the symmetry of the potential in the strong confinement limit, the solution of the Schrodinger Eq. for positive energies are two-fold degenerate with states of even and odd parity represented by $\eta_{even}$ and $\eta_{odd}$,
\begin{equation}\label{TransverseExcitedEven}
\eta_\nu^{even}(\tau) = -\sqrt{\frac{2}{a}}\cos(\nu\pi) \sin \left( \frac{2 \nu \pi }{a}|\tau| \right)
\end{equation}
and
\begin{equation}\label{TransverseExcitedOdd}
\eta_\nu^{odd}(\tau) = -\sqrt{\frac{2}{a}} \cos(\nu\pi) \sin \left( \frac{2 \nu \pi}{a} \tau \right) ,
\end{equation}
where $\nu$ takes on any positive integer value. It is straightforward to show that the wavefunctions in Eqs. (\ref{TransverseGround}), (\ref{TransverseExcitedEven}) and (\ref{TransverseExcitedOdd}) are orthogonal when $g \rightarrow \infty$. Thus the contribution of the transverse component to the energy for $E_\tau >0$ is given by
\begin{equation}\label{TranverseEnergy}
E_\nu^\tau = \frac{2 \nu^2 \pi^2 \hbar^2}{m a^2}.
\end{equation}
Fig. \ref{fig:potential} shows the transverse ground state and the two lowest-energy ``continuum" states.

The even and odd parity wavefunctions in the extreme confinement limit given by Eqs. (\ref{TransverseExcitedEven}) and (\ref{TransverseExcitedOdd}) corresponding to the same state index $\nu$ have the same eigenenergies. Since the transition moment between wavefunctions with the same parity vanishes, the only nonzero transition moments from the ground state, an even parity state, are to odd parity excited states.

Note that from this point on, we differentiate between even and odd parity states using positive and negative integers, so that even(odd) wavefunctions are represented by positive (negative) integers. For example $y_{-n,m}$ is a transition between the $n^{th}$ odd-parity state, given by Eq. (\ref{TransverseExcitedOdd}), and $m^{th}$ even-parity states, given by Eq. (\ref{TransverseExcitedEven}). The ground state is not degenerate and caries the index $0$..

\section{Sum rules}

The Thomas-Reiche-Kuhn sum rules are calculated using the commutator between the position operator and the Hamiltonian of the quantum system, and relate the position matrix elements, $x_{ij}$, and the energies, $E_i$, $E_j$, \ldots to each other. For a one-dimensional system, the sum rules are given by
\begin{eqnarray}\label{sumRules}
\sum_n^{\begin{array}{c}
          \mbox{all} \\
          \mbox{states}
        \end{array}} \left(E_{n0} - \frac{E_{m0} + E_{p0}}{2}\right) x_{mn}x_{np} &=& \frac{\hbar^2 N}{2m} \delta_{mp}, \nonumber \\
\end{eqnarray}
where $m$, $n$, and $p$ are labels of the three eigenstates of the system with energies $E_m$, $E_n$, and $E_p$,  the ground state is denoted by $0$, $E_{i0} = E_i - E_0$, $N$ is the number of electrons and $\delta_{mp}$ is the Kronecker delta function. Throughout this text, we loosely call $x_{ij}$ the transition moment between states $i$ and $j$. The summation spans the complete set of eigenstates of the system, including both degenerate and non-degenerate states\cite{fernández2002thomas} as well as discrete and continuum states \cite{Bethe77.01}.

Sum rules are used in many areas of physics.\cite{belloni08.01} In nonlinear optics, they are used to calculate the fundamental limits of the first and second hyperpolarizabilities, $\beta$\cite{kuzyk00.01} and $\gamma$\cite{kuzyk00.01, kuzyk00.02}, to find a dipole-free sum over states expression for calculating nonlinear hyperpolarizabilities\cite{kuzyk05.02, perez01.08}, and to define constraints on the transition moments and energies in Monte Carlo simulations of the first\cite{kuzyk08.01} and second\cite{shafei10.01} hyperpolarizabilities. While the verification of sum rules in one dimension is a classic problem in quantum mechanics\cite{cohen1977quantum}, there are few studies of lower-dimensional systems that are embedded in a higher-dimensional space.

Not only can the sum rules be employed in the study of the upper bounds of the nonlinear response of a system of quantum wires, they can also be used to test the validity of the solutions to Schrodinger Equation. For example, as we will show below, the sum rules appear to be violated for \emph{ideal} quantum wires, but come into compliance when \emph{leaky} quantum wires with tunneling\cite{exner2007leaky} are evaluated in the limit of full confinement.

Consider a quantum wire with total length $L$, as shown in Fig. \ref{fig:2Dwire}. Along the $s$ axis the problem reduces to one dimension, so the sum rules can be written as,
\begin{equation}\label{SRs}
\sum_n^\infty \left(E_{n0} - \frac{E_{m0} + E_{p0}}{2}\right) s_{mn}s_{np} = \frac{\hbar^2 N}{2 m} \delta_{mp}
\end{equation}
where $s$ is the distance along the wire from its end at the origin to the coordinates $(x,y)$ and $s_{mn} = \left<m\left|s\right|n\right>$. To verify the sum rules in the $x$ direction, and ignoring the transverse contribution, i.e. $\tau = 0$ under extreme confinement, we substitute $x = s \cos\theta$ into Eq. (\ref{SRs}), where $\theta$ is the angle between the wire and the $x$ axis, yielding
\begin{equation}\label{SRx}
\sum_n^\infty \left(E_{n0} - \frac{E_{m0} + E_{p0}}{2}\right) x_{mn}x_{np} = \frac{\hbar^2 N}{2 m}\cos\theta^2 \delta_{mp}.
\end{equation}
Equation (\ref{SRx}) is in apparent contradiction to Eq. (\ref{sumRules}).  Later, we will see that this contradiction can be resolved by including the transverse infinite-energy continuum states.

It is instructive to consider a similar apparent violation of the sum rules studied by Hadjimichael \emph{et al.} for a rigid rotator in three dimensions,  with Hamiltonian,
\begin{equation}\label{RigidRotator}
H = \frac{1}{2\mu R^2}\left(p^2_\theta + \frac{p^2_\phi}{\sin^2\theta}\right) ,
\end{equation}
where, $p_i$ is the momentum of the particle along the $i^{th}$ coordinate, $\mu$ is the reduced mass and $R$ is the radius of the particle's path.\cite{hadji97} They argue that in the classical picture, the radius of the rigid rotator is fixed, which implies zero momentum in the radial direction, or $p_r = 0$. In the quantum interpretation, the uncertainty principle for transverse confinement with $\Delta R \simeq 0$ demands that the radial component of the momentum must tend to infinity. Therefore, these high-momentum states (and therefore also of high energy) contribute to the transverse transition moments in the radial direction.

To model this system, they used an attractive radial delta function potential, $V(r) = -g\delta(r-R)$ with $g>0$ as the potential strength. This system resembles the rigid rotator when the particle is in its {\em transverse} ground state when $g \rightarrow \infty$ - the extreme confinement condition. Hadjimichael \emph{et al.} used this approach to verify numerically that the sum rules are obeyed for the rigid rotator in the limit when the transverse energies tend to infinity.

The uncertainty principle imposes a similar constraint on the transverse component of the electron's wavefunction in a quantum wire, leading to infinite energy states that contribute to sum rules. This suggests that \emph{leaky quantum wires},\cite{exner2007leaky} with continuum states included must be used in any realistic model of a quantum wire.

In the spirit of this approach, we verify the sum rules for the one-dimensional transverse wavefunction of a quantum wire as a limiting case of a leaky wire under strong confinement. Then, taking into account both longitudinal and transverse wavefunctions, we verify that the sum rules of a single quantum wire in two-dimensional space is obeyed when continuum states of infinite energy and infinitesimal transition moment are included.

\subsection{Sum rules for transverse direction of a quantum wire}

Belloni and Robinett have verified the sum rules for a single delta function potential.\cite{belloni08.01} Here we verify the sum rules for a system with a delta function potential in a box, where the states of the box are used to approximate the transverse wavefunction of a quantum wire in the limit when the box becomes large. Based on our convention in the preceding sections, we use Latin and Greek characters for longitudinal and transverse wavefunctions, respectively. In this section, however, we will reformulate the problem by projecting all quantities along the $y$ axes, thus eliminating the variable $\tau$, and replacing $\nu$ with $n$.  We also drop the $\tau$ superscript for simplicity.

Considering the ground state sum rules, i.e. $(m,p) =(0,0)$ in Eq. (\ref{sumRules}), we need to calculate the moment $\left<0\left|y\right|n\right> \equiv y_{0n}$ where the subscripts $0$ and $n$ denote the transverse ground and $n^{th}$ excited states, respectively. The transition moment, $y_{0n}$, is determined using the inner product of the ground state wavefunction, Eq. (\ref{TransverseGround}), with the odd-parity excited state wavefunctions according to Eq. (\ref{TransverseExcitedOdd}). We find,
\begin{eqnarray}\label{ymomentsLimit}
y_{-n,0} &=& \int_{-a/2}^{a/2} \eta_n^{odd}(y) y \eta_0(y)\nonumber \\
&=& \frac{2^{7/2} (-1)^{n+1} \pi n a^{5/2} k_0^{3/2}}{\left(a^2 k_0^2 + 4 \pi^2 n^2\right)^2}.
\end{eqnarray}
Using the energy difference
\begin{equation}\label{EnergyDiffEn0}
E_{n0} = \frac {\hbar^2} {2m}\left(k_n^2-k_0^2\right),
\end{equation}
and Eq. (\ref{ymomentsLimit}), we find
\begin{eqnarray}\label{SumRuleTransverseY}
&&\lim_{k_{0}\rightarrow \infty} \sum_n \left| y_{-n,0} \right|^2 E_{n0} = \lim_{k_{0}\rightarrow \infty} \int dn \left| y_{-n,0} \right|^2 E_{n0}\nonumber \\
&=&\lim_{k_{0}\rightarrow \infty} \int dn \frac{64 a^3 \pi ^2 k_0^3 n^2 \hbar ^2}{m \left(a^2 k_0^2 + 4 \pi^2 n^2\right)^3} = \frac {\hbar^2} {2m}
\end{eqnarray}
where according to Eq. (\ref{definektau0}), $k_{0}\rightarrow \infty$ corresponds to $g\rightarrow \infty$.

To verify the diagonal sum rules for the transverse odd-parity first excited state, i.e. $(m,p)=(-1,-1)$ in Eq. (\ref{sumRules}) in the $y$ direction, we start by calculating $y_{n,-1}$,
\begin{eqnarray}\label{yn1}
y_{n,-1} = \int_{-a/2}^{a/2} \eta_n^{even}(y) y \eta_1^{odd}(y) = \frac{2 a n(1+(-1)^n)}{\left(n^2-1\right)^2 \pi^2}.  \nonumber \\
\end{eqnarray}
Using Eqs. (\ref{EnergyDiffEn0}) and (\ref{yn1}) we get,
 \begin{equation}\label{yn1en1}
\left|y_{n,-1}\right|^2 E_{n,1} = \frac{8\left(1+ (-1)^n\right)^2 n^2 \hbar^2}{\left(n^2 -1\right)^2 \pi^2}.
\end{equation}
Summing over all even wavefunctions yields
\begin{eqnarray}\label{SR(1,1)}
\lim_{k_{0}\rightarrow \infty} \int dn \left|y_{n,-1}\right|^2 E_{n1} &=& \lim_{k_{0}\rightarrow \infty} \int dn \frac{128 n^2 \hbar^2}{m(4 n^2 - 1)^2 \pi^2}\nonumber \\
&=& \frac{\hbar^2}{2 m}
\end{eqnarray}

For the $(m,p)=(-2,-2)$ diagonal sum rule, we need $y_{n,-2}$, which is obtained as is $y_{n,-1}$ above, yielding
\begin{equation}\label{yn2}
y_{n,-2} = \frac{4 (-1)^n (-1 + (-1)^n) a n}{(-4 + n^2)^2 \pi^2} .
\end{equation}
Summing over all even wavefunctions yields,
\begin{eqnarray}\label{yn2en2}
&& \lim_{k_{0}\rightarrow \infty} \sum_n^{\infty} \left|y_{n,-2}\right|^2 E_{n2}= \nonumber \\
&&\lim_{k_{0}\rightarrow \infty} \int dn \frac{32\left(-1 + (-1)^n\right)^2 n^2 \hbar^2}{m \pi^2 \left(n^2 - 4\right)^3} = \frac{\hbar^2}{2 m}.
\end{eqnarray}
This calculation can be repeated for each diagonal sum rule, leading to verification of the sum rules for all odd-parity eigenstates. The same procedure can be followed for even-parity eigenstates.

For the non-diagonal sum rules, ($m \neq p$ in Eq. (\ref{sumRules})), we use the fact that symmetry demands that the transition moment between any two states of the same parity vanishes. Thus in expressions such as $y_{mn} y_{np}$, the parity of states $m$ and $p$ must be the same but opposite in parity to state $n$. Using this constraint to calculate $y_{mn}$ and $y_{np}$, the non-diagonal sum rules, i.e. when $m \neq p$, become
\begin{eqnarray}
 \sum_n^\infty \left(E_{nm} + E_{np}\right) y_{mn} y_{np} &=& -\frac{m p \hbar^2\left(i^{2m} - i^{2p}\right)}{m\left(m^2 - p^2\right)} = 0 , \nonumber \\
\end{eqnarray}
where $i^{2} = -1$. The above result holds for any arbitrary odd wavefunction. Verification of non-diagonal sum rules for even-parity eigenstates proceeds along the same lines, except that transition moments from an even-parity state to the ground state vanish because the ground state is of even parity. Hence, the sum rules are verified for the delta potential inside a box potential.

\subsection{Sum rules for a quantum wire in 2-D space}

The longitudinal, $s$, and transverse coordinate, $\tau$, are related to the fixed axes $x$ and $y$ through the rotation matrix according to,
\begin{eqnarray}\label{stauxy}
x = s \cos \theta \mp \tau \sin\theta, \nonumber \\
\\
y = s \sin \theta \pm \tau \cos\theta, \nonumber
\end{eqnarray}
where the upper (lower) sign is for the case when the electron is above (below) the wire. The transition moments follow,
\begin{eqnarray}\label{xTransitions}
x_{mp} &=& \left< m \left|x\right|p\right> = \cos\theta \left< m \left|s\right|p\right> \mp \sin\theta\left< m \left|\tau\right|p\right> \nonumber \\
&=& s_{mp} \cos\theta \mp \tau_{mp} \sin\theta.
\end{eqnarray}
whence,
\begin{eqnarray}
x_{mn}x_{np} &=& s_{mn}s_{np}\cos^2\theta + \tau_{mn}\tau_{np}\sin^2\theta  \nonumber \\
&& \mp\sin\theta \cos\theta \left(s_{mn}\tau_{np} + s_{np}\tau_{mn}\right).
\end{eqnarray}
Then, the sum rules are given by,
\begin{eqnarray}\label{xSR}
&&\sum_n\left(E_n - \frac{E_m + E_p}{2}\right) x_{mn}x_{np} = \frac{\hbar^2}{2m} \delta_{mp} \times \nonumber \\
&& \left(\sin^2\theta + \cos^2\theta\right) \mp \sin\theta \cos\theta \times\nonumber \\
&& \sum_n\left(E_n - \frac{E_m + E_p}{2}\right) \left(s_{mn}\tau_{np} + s_{np}\tau_{mn}\right) \nonumber \\
&=& \frac{\hbar^2}{2m} \delta_{mp}
\end{eqnarray}
where we have used the fact that the sum rules are obeyed in the $s$ and $\tau$ directions, and the fact that
\begin{eqnarray}\label{mixedSR}
&& \sum_n\left(E_n - \frac{E_m + E_p}{2}\right) \left(s_{mn}\tau_{np} + s_{np}\tau_{mn}\right)\nonumber \\
&=& \frac{1}{2} \left<m\left|\left[s, \left[H, \tau\right]\right]\right|p\right>  + \frac{1}{2} \left<m\left|\left[\tau, \left[H,s\right]\right]\right|p\right> \nonumber \\
&=& \frac{1}{2} \left<m\left|\left[s, \tau\right]\right|p\right>  + \frac{1}{2} \left<m\left|\left[\tau, s\right]\right|p\right> = 0,
\end{eqnarray}
since both terms in the last line of Eq. (\ref{mixedSR}) are zero.

\section{Linear and Nonlinear Response}

The central result of this work is that the transverse contribution to the NLO properties of a single quantum wire in the strong confinement limit is negligible. To show this, we begin by calculating the contribution of the transverse wavefunction to the polarizability, $\alpha$, hyperpolarizability, $\beta$, and the second hyperpolarizability, $\gamma$, in the off-resonant regime.

In the tightly-confined limit, the polarizability projected onto the y axis is given by
\begin{equation}\label{alpha}
\alpha = 2 e^2 {\sum_n}' \frac{y_{0n}y_{n0}}{E_{n0}}
\end{equation}
where the prime indicates that the ground state is excluded from the summation. Using Eqs. (\ref{ymomentsLimit}) and (\ref{EnergyDiffEn0}) for the transverse contribution to $y_{n0}$ and $E_{n0}$, and using the fact that $y_{0n} = y_{n0}^*$, Eq. (\ref{alpha}) gives,
\begin{eqnarray}\label{alpha-2}
\alpha &=& 2e^2 \lim_{k_0 \rightarrow \infty}{\sum_{n}}'\frac{\left|y_{n0}\right|^2}{E_{n0}} \simeq \lim_{k_0 \rightarrow \infty}\int_{-\infty}^0 \frac{\left|y_{-n,0}\right|^2}{E_{n0}} dn \nonumber \\
&=&\lim_{k_0 \rightarrow \infty} \frac{5 m}{8 k_0^4 \hbar ^2} = 0 .
\end{eqnarray}
Thus, in the full-confinement limit ($g\rightarrow\infty$), $\alpha$ approaches zero since $k_0 \propto g$. Hence, the transverse wavefunction of a quantum wire does not contribute to the linear response.

The second-order polarizability, $\beta_{yyy}\equiv \beta$ is given by
\begin{eqnarray}\label{beta}
\beta = -3e^3 {\sum_{m=-\infty}^{\infty}}'{\sum_{n = -\infty}^{\infty}}' \frac{y_{0n}\bar{y}_{nm}y_{m0}}{E_{n0}E_{m0}} ,
\end{eqnarray}
where $\bar{y}_{nm} = y_{nm} - y_{00}\delta_{nm}$. Since the ground state has even parity, states $m$ and $n$ must have odd parity to yield nonzero $y_{0m}$ and $y_{n0}$. However under these conditions, $y_{mn}$ vanishes because states $m$ and $n$ have the same parity. So the transverse wavefunction does not contribute to the first hyperpolarizability.  This result also follows from the fact that the hyperpolarizability for a symmetric potential vanishes.

The non-resonant second hyperpolarizability, $\gamma_{yyyy} \equiv \gamma$, with $y_{00}=0$ is given by
\begin{eqnarray}\label{gamma}
\gamma &=& 4e^4\left({\sum_{n=-\infty}^{\infty}}'{\sum_{p = -\infty}^{\infty}}'{\sum_{q=-\infty}^{\infty}}'\frac{y_{0n}y_{np}y_{pq}y_{q0}}{E_{n0}E_{p0}E_{q0}} \right. \nonumber \\
&&\left.  - {\sum_{n=-\infty}^{\infty}}'{\sum_{p = -\infty}^{\infty}}'\frac{|y_{0n}|^2 |y_{0p}|^2}{E_{n0}^2E_{p0}}\right).
\end{eqnarray}
For the first summation not to vanish, states $n$ and $q$ must have odd parity and state $p$ must be of even parity. In the second summation, states $n$ and $p$ both must have odd parity. The first summation in Eq. (\ref{gamma}) can be written as
\begin{eqnarray}\label{firstSummation}
{\sum_{n,p,q}}' &=& {\sum_{|n|\neq \left|p\right| \neq |q|}}' + {\sum_{|n|= \left|p\right|\neq |q|}}' + {\sum_{|n|\neq  \left|p\right| = |q|}}' +\nonumber \\ &&{\sum_{|n|=|q|\neq  \left|p\right|}}' + {\sum_{|n|= \left|p\right|=|q|}}'
\end{eqnarray}
The last term of Eq. (\ref{firstSummation}), when $|n|= \left|p \right|=|q|$, yields,
\begin{eqnarray}
&&\lim_{k_0 \rightarrow \infty}{\sum_{n}}'  \frac{\left|y_{0,-n}\right|^2 \left|y_{n,-n}\right|^2}{E_{n0}^3} \nonumber \\
&\simeq& \lim_{k_0 \rightarrow \infty} \int dn \frac{64 a^{13} k_0^3 m^3 \pi^2 n^2}{\left(a^2 k_0^2 + 4 \pi^2 n^2\right)^7 \hbar^6} \rightarrow 0 ,
\end{eqnarray}
where we have used the fact that $y_{n,-n} = a/4$. It is straightforward to verify that all other terms in Eq. (\ref{firstSummation}) vanish in the limit of full confinement, as calculated in the Appendix.

We can express the second term in Eq. (\ref{gamma}) in the form
\begin{eqnarray}\label{gamma2ndTerm}
&&\lim_{k_0 \rightarrow \infty}  \sum_{n,p}^\infty\frac{|y_{0,-n}|^2 |y_{0,-p}|^2}{E_{n0}^2E_{p0}} = \lim_{k_0 \rightarrow \infty} \int dn \, dp \times\nonumber \\
&& \frac{131072 a^{16} k_0^6 m^3 \pi^4 n^2 p^2}{\left(a^2 k_0^2 + 4 \pi^2 p^2\right)^5 \left(a^2 k_0^2 + 4 \pi^2 n^2\right)^6 \hbar^6}\rightarrow 0 ,
\end{eqnarray}

Consequently, the contribution of the transverse wavefunction in the full-confinement limit to the linear and nonlinear response of a quantum wire is zero.

\section{Conclusion and summary}

The present work lays the foundations for studies of geometrical effects on the NLO response of a complex system such as a network of quantum wires, which can be assembled into various configurations such as quantum loops, bent wires, and more generally, quantum graphs.  We have verified that the transverse sum rules are obeyed in the highly-confined limit only when the infinite-energy continuum states are included.  The longitudinal sum rules are trivially obeyed in a quantum wire, so our results are general in that they show that the full sum rules must be obeyed for any wire orientation and thus any collection of connected segments.

Our work resolves an apparent paradox of the semiclassical view of confinement.  By using a limiting procedure, we show that the infinite-energy continuum states can both contribute to the sum rules while also explaining how they do not contribute to the linear and nonlinear susceptibilities.  For these states, in the strong confinement limit, the transition moments tend to zero and the energies tend to infinity.  Thus, the susceptibilities, which are of the form $x^{n+1}/E^n$, will tend to zero while the individual terms that contribute to the sum rules are of the form, $x^2 E$, which are finite, and nonzero.  Thus, the classical picture of confinement can be used to calculate the nonlinear response.

To summarize,
\begin{enumerate}
  \item Under extreme quantum confinement, the transverse nonlinear response vanishes.
  \item The sum rules remain valid even for special cases such as quantum wires in the tight-confinement limit.
  \item Being derived from the sum rules, the fundamental limits of the nonlinear susceptibility remains unchanged for lower-dimensional systems.
  \item The nonlinear susceptibility of a reduced-dimensional system will be lowered when the infinite-energy transverse states are ignored.
  \item The classical picture of the idealized quantum wire miss the infinite-energy transverse-confined states, thus the sum rules appear to be violated. Neglect of these transverse states, however, has no effect on the nonlinear susceptibilities.
\end{enumerate}

\begin{acknowledgments}
We would like to thank the National Science Foundation (NSF) (ECCS-0756936) and Wright Paterson Air Force Base for generously supporting this work.
\end{acknowledgments}

\appendix
\section{Calculation of $\gamma$ terms}\label{app:appendix}
In this appendix, we evaluate each term in  Eq. (\ref{firstSummation}) to show that they all vanish.  Note that all analytical expressions were evaluated using Mathematica.\textregistered

The first through fourth term in Eq. (\ref{firstSummation}) yield,
\begin{widetext}
\begin{eqnarray}
&&\lim_{k_0 \rightarrow \infty}{\sum_{|n|\neq |p|\neq |q|}}'\frac{y_{0,-n}y_{-n,p}y_{p,-q}y_{-q,0}}{E_{n0}E_{p0}E_{q0}} \nonumber \\
&=& \lim_{k_0 \rightarrow \infty} \sum_{|n|\neq |p|\neq |q|} \frac{4096 (-1)^{q + n} \left(-1 + (-1)^{n+p}\right) \left(-1 + (-1)^{n+q}\right) a^{13} k_0^3 m^3 n^3 p^2 q}{\pi^2 \left(a^2 k_0^2 + 4 \pi^2 p^2\right)^3 \left(a^2 k_0^2 + 4 \pi^2 q^2\right) \left(q^2 - n^2\right)^2 \left(n^2 - p^2\right)^2 \left(a^2 k_0^2 + 4 \pi^2 n^2\right)^3 \hbar^6} = 0,
\end{eqnarray}
\end{widetext}
where we test the convergence using a summation that spans 25 states. The second and third term on the right-hand side of Eq. (\ref{firstSummation}) are identical. The second term becomes
\begin{widetext}
\begin{eqnarray}
\lim_{k_0 \rightarrow \infty}{\sum_{|n| = |p|\neq |q|}}' \frac{y_{0,-n}y_{-n,p}y_{p,-q}y_{-q,0}}{E_{n0}E_{p0}E_{q0}} &=&  \lim_{k_0 \rightarrow \infty}{\sum_{|n|\neq |q|}}\frac{-512 \left(-1 + (-1)^{n+q}\right) a^{13} k_0^3 m^3 q^2 n^2}{\left(a^2 k_0^2 +  4 \pi^2 m^2\right)^3 \left(q^2 - n^2\right)^2 \left(a^2 k_0^2 + 4 \pi^2 n^2\right)^4 \hbar^6} = 0
\end{eqnarray}
\end{widetext}
and finally the fourth term in Eq. (\ref{firstSummation}) yields,
\begin{widetext}
\begin{eqnarray}
\lim_{k_0 \rightarrow \infty}{\sum_{|n| = |q| \neq |p|}}' \frac{y_{0,-n}y_{-n,p}y_{p,-q}y_{-q,0}}{E_{n0}E_{p0}E_{q0}} &=& \lim_{k_0 \rightarrow \infty} {\sum_{|n|\neq |p|}}' \frac{4096 \left(-1 + (-1)^{n+p}\right)^2 a^{13} k_0^3 m^3 p^2 n^4}{\pi^2 \left(a^2 k_0^2 + 4 \pi^2 p^2 \right) \left(p^2 - n^2\right)^4 \left(a^2 k_0^2 + 4 \pi^2 n^2\right)^6 \hbar^6} = 0.
\end{eqnarray}
\end{widetext}

\end{document}